**Short communication**

# Expanding the phenotype of SCA19/22: Parkinsonism, cognitive impairment and epilepsy.

Vincent Huin [a, b, *], Isabelle Strubi-Vuillaume [b], Kathy Dujardin [c, d], Marine Brion [d], Marie Delliaux [d], Delphine Dellacherie [e], Jean-Christophe Cuvellier [e], Jean-Marie Cuisset [e], Audrey Riquet [e], Caroline Moreau [c, d], Luc Defebvre [c, d], Bernard Sablonnière [a, b], David Devos [c, d, f]

[a] Univ. Lille, Inserm, CHU Lille, UMR-S 1172 - JPArc - Centre de Recherche Jean-Pierre AUBERT Neurosciences et Cancer, F-59000 Lille, France

[b] CHU Lille, Institut de Biochimie et Biologie moléculaire, Centre de Biologie Pathologie et Génétique, F-59000 Lille, France

[c] Univ. Lille, Inserm, CHU Lille, U1171 - Degenerative & vascular cognitive disorders, F-59000 Lille, France

[d] CHU Lille, Service de Neurologie et Pathologies du Mouvement, Hôpital Roger Salengro, F-59000 Lille, France

[e] CHU Lille, Service de Neuropédiatrie, Hôpital Roger Salengro, F-59000 Lille, France

[f] CHU Lille, Service de Pharmacologie médicale, Faculté de Médecine, F-59045 Lille, France

*Corresponding author: Vincent Huin

Inserm UMR-S 1172, JPArc, rue Polonovski, F-59045, Lille, France

Tel: +33 320622075

Fax: +33 320538562

vincent.huin@inserm.fr

**Running title**

New phenotype in SCA19/22

## ABSTRACT

**Introduction:** Spinocerebellar ataxia types 19 and 22 (SCA19/22) are rare conditions in which relatively isolated cerebellar involvement is frequently associated with cognitive impairment. Here, we report on new clinical features and provide details of the cognitive profile in two SCA19/22 families.

**Methods:** Two families displaying an autosomal-dominant form of cerebellar ataxia underwent clinical examinations and genetic testing.

**Results:** In addition to the classical clinical features of SCA, a wide spectrum of cognitive disorders (including visuospatial impairments) was observed. Eight patients had mild Parkinsonism, and five had epilepsy. Genetic testing showed that the *KCND3* mutation (c.679_681delTTC, p.F227del) was present in both families.

**Conclusions:** Our findings broaden the phenotypic spectrum of SCA19/22, and suggest that *KCND3* should be included in the list of candidate genes for epilepsy, Parkinsonism and cognitive impairment.

## 1. Introduction

Spinocerebellar ataxia types 19 and 22 (SCA19/22) are, rare inherited neurodegenerative disorders characterized by slowly progressing ataxia, frequent cognitive impairment, and signs of frontal lobe dysfunction [1,2]. Other neurological features observed in some (but not all) cases include postural head tremor, myoclonus, pyramidal signs and neuropathy [3]. SCA19 and SCA22 almost certainly correspond to the same disease, which is caused by loss-of-function mutations within the potassium voltage-gated channel subfamily D member 3 (*KCND3*) gene [4]. The *KCND3* gene encodes the voltage-gated potassium channel Kv4.3, which is known to be important for repolarization of the action potential in excitable cells [5]. In SCA19/22, mutations alter the channel's location and/or function. In turn, this alters the excitability of Purkinje neurons and leads to their neurodegeneration [6]. Here, we describe two novel SCA19/22 families and broaden the phenotypic spectrum to encompass mild Parkinsonism and epilepsy.

## 2. Methods

### *2.1. Clinical examination*

Two unrelated French families displaying ataxia with an autosomal-dominant inheritance pattern were examined in the Department of Neurology at Lille University Medical Center (Fig. 1). All affected family members underwent a detailed clinical evaluation. During examination, Parkinsonism was defined with the classical association of rigidity and akinesia that can be associated with a rest tremor. The symptoms of four deceased family members were also recorded. Repeat expansions in the SCA1, 2, 3, 6, 7, 8, 12, 17 genes and conventional mutations in the SCA13, 14, 21, 28 genes were ruled out. All individuals gave their written informed consent to participation in the study.

### 2.1. Cognitive and behavioral measures

Twelve affected individuals underwent a neuropsychological evaluation. Given that some individuals had trouble reading and writing (due to learning disabilities and ataxia), an adapted testing procedure was mainly focused on non-verbal tests. Overall cognition was assessed with the Montreal Cognitive Assessment (MoCA). Raven's colored progressive matrices (PM47) were used to assess abstract reasoning and non-verbal intelligence. Attention and working memory were assessed with the forward and backward Corsi block-tapping test. We used simple and choice reaction time tests to assess vigilance, inhibition and divided attention. Visuospatial learning and memory performance were tested with the 10/36 spatial recall test. Executive function was analyzed with the Tower of London test, with four difficulty levels: (i) “3N” items, where the goal can be achieved in a minimum of 3 moves, (ii) “5N” items, where the goal can be achieved with a minimum of 5 moves, and where initially no bead can be brought directly to its final destination, (iii) “5Iþ” items, where the goal can be achieved with a minimum of 5moves, and one bead can be immediately moved into its final destination, (iv) “5I-” items, where the goal can be achieved with a minimum of 5 moves, and one bead can be immediately moved into its final destination but will immediately prevent the solution of the problem. Performance was assessed by the mean number of moves at each difficulty level. Given the individuals' ataxia, the test was not timed. A 15-item version of the Benton judgment of line orientation (BJLO) test was used to assess visuospatial abilities. Receptive language abilities were rated with the Token test. The Neuropsychiatric

Inventory was administered to determine the presence and severity of behavioral disorders.

### *2.3. Molecular analysis*

Genomic DNA was isolated from peripheral blood, according to standard procedures. Targeted, next-generation sequencing of 24 genes involved in dominant hereditary ataxia was performed in the probands and in one healthy sibling (Data S1). Variations were confirmed by Sanger sequencing. Repeat expansion in the SCA10 gene was ruled out using a fluorescent repeat-primed PCR assay [7].

## 3. Results

### *3.1. Clinical features*

The probands were referred to our clinic for gait disorders, poor balance and dysarthria. The family medical history was suggestive of an inherited disease with autosomal-dominant transmission. Table 1A summarizes the clinical characteristics of the 16 affected patients.

Eleven patients (68.8% of the carriers) developed a classic SCA19/22 phenotype, i.e. slowly progressing cerebellar ataxia with predominant gait impairment (Video 1 and 2). Poor balance was the first symptom to be identified. As the disease progressed, the patients developed gait disorders, dysarthria, nystagmus and then dysphagia. The estimated mean age at ataxia onset was 23.1 years (range: 2e66). Mild Parkinsonism was observed in eight patients (50% of the carriers) (III-6, III-9, IV-5, IV-6, IV-9, V-2 in family A and II-2, III-2 in family B).

The phenotype was dominated by epilepsy in five younger patients (31.3%). In family A, the patients IV-10, V-2, V-4 and V-6 suffer from epilepsy with a mean age at epilepsy onset of 5.3 years (range: 3e12), whereas ataxia was mild, delayed or not noticed. The patient IV-10 had a history of tonic-clonic epilepsy between the age of 3 and 6. The patient V-2 had complex partial seizures sometimes complicated by generalized tonic-clonic seizures. His brother (V-4) had complex partial seizures (temporal area) followed by generalized atonic seizures. He experienced one episode of tonic-clonic seizure. The patient V-6 had complex partial seizure corresponding to benign rolandic epilepsy and complicated by generalized atonic seizures. EEG in patients V-2 and V-4 revealed spike wave or polyspikewave discharges in the right frontotemporal area (V-2) or in the right frontal area (V-4). EEG in patient V-6 revealed paroxysmal rhythmic theta waves or biphasic theta waves in the right frontotemporal region. The fifth and last patient (IV-1 in family B) reported frequent myoclonus. The EEG of patients V-4 and V-6 in family A are depicted in supplementary Figure S2.

Brain MRI datasets were available for seven patients and showed isolated atrophy of the vermis in the five oldest individuals (III-9, IV-1, IV-6, IV-7 in family A and II-2 in family B).

### *3.2. Cognition, behavior and neuropsychiatric disorders*

The results of the comprehensive neuropsychological test battery are shown in Table 1B. Most patients had experienced learning difficulties, and five (31.3%) had attended special schools. For affected adults, the time spent in formal education ranged from 8 to 14 years (median: 10 years). The estimated intelligence quotient was >80 for ten patients (83.3%) and between 60 and 70 for two patients (16.7%).

Impairments in overall cognition were detected in all but two of the twelve individuals having undergone neuropsychological testing. However, this impairment was severe (MoCA score <21) in only four individuals. Most individuals achieved normal levels of performance in the attention and working memory tests. Although levels of performance in the visuospatial learning and memory tests were normal for immediate recall, most subjects had impairments for delayed recall. Half of the individuals displayed impairments in visuospatial functions. Planning and inhibition abilities were relatively unaffected in most individuals. In two of the younger patients in family A (V-4 and V-6), cognitive impairment had been suspected during clinical examination, and both had learning delay. One patient (I-1 in family B) died before cognitive performance could be assessed. In summary, 12 of the 16 patients (75%) in whom cognitive function had been assessed were found to display mild cognitive impairment.

Behavioral disorders were detected in four patients: slight aggressiveness (n = 2), mild agitation (n = 1), mild depression (n = 3), severe apathy (n = 1), and severe sleep disorders (n = 1). A history of major depressive disorder was reported by five patients, and obsessive-compulsive disorder was reported by one. Two patients were taking medication for anxiety.

### ***3.3. Genetic testing***

Next-generation sequencing revealed the same *KCND3* mutation (c.679_681delTTC, p.F227del) (NM_004980.4) in the two families. Sanger sequencing confirmed the segregation of this mutation in all affected individuals and one unaffected individuals (Supplemental Figure S1). The ATTCT pentanucleotide repeat expansion in the intron 9 of *ATXN10* (SCA10) was in the normal range with all patients having between 10 and 18 ATTCT repeats.

## 4. Discussion

The F227del variation is a recurrent mutation that has been previously reported in different ethnic groups, e.g. in a Han Chinese family and a French family with SCA19/22 diagnosis [2]. This mutation is an in-frame deletion of a highly conserved amino acid residue in KCND3 that reduces the protein length and is absent from controls in Exome Sequencinq Project, 1000 Genomes Project, Exome Aggregation Consortium and the Genome Aggregation Database. In vitro functional studies has been performed in two previous studies [2,5] and are supportive of a damaging effect with defect in the trafficking and cell surface expression of Kv4.3 and thus loss of the potassium channel current amplitude.

In the SCA19/22 patients assessed here, Parkinsonism did not have an impact on daily living and did not justify treatment with Ldopa, since this might have worsened balance even more. However, the presence of Parkinsonism might help the clinician to diagnose SCA19/22 because this sign is not frequent in most forms of SCA [8]. Parkinsonism has already been described in SCA2, 17, 21 and also in SCA3 or Machado-Joseph disease that is the most common SCA worldwide. Parkinsonism in in SCA3, especially the akinetic-rigid syndrome is considered as a subphenotype of the SCA3 and may respond to treatment by levodopa or dopamine agonists. The existence of Parkinsonism features in SCA3 is not fully understood but could involve dysfunction of the midbrain dopaminergic system caused by neurodegeneration of the basal ganglia and especially of the substantia nigra [9].

Interestingly, mild cogwheel rigidity was noted in two patients (aged 61 and 77) with an F227del mutation in the French family described by Lee et al. [2]. Moreover, extrapyramidal symptoms are compatible with (i) the known expression pattern for Kv4.3, and (ii) the moderate neurodegeneration of the substantia nigra reported in a neuropathological examination of a SC19/22 patient who suffered from isolated ataxia and mild cognitive impairment [10]. Indeed, neurodegeneration in SCA19/22 is not restricted to the cerebellum alone but also involves distinct brainstem nuclei (such as the substantia nigra, the raphe nucleus and the inferior olivary nucleus).

Schelhaas et al. reported that most SCA19 patients had cognitive impairment (including executive function) and mood disturbance, despite a MMSE score >24. They also observed a low premorbid IQ (below 105) [3,4]. Most of the individuals assessed in the present study also had a rather low IQ. However, when assessed with the MoCA (which is more sensitive to impaired executive function), overall cognitive efficiency was impaired in most individuals. We also observed the frequent impairment of visuospatial function. The performance pattern for visuospatial memory revealed a frequent impairment in the long-term storage of information via episodic memory. Furthermore, mood disturbances and behavioral disorders were frequent in the two families studied here. Overall, this pattern of cognitive deficits and behavioral disorders is close to the cerebellar cognitive affective syndrome as introduced by Schahmann and Sherman [11]. This syndrome due to damage to the cerebellum is characterized by the association of deficits in executive function, visuospatial cognition and language with behavioral disorders, namely impulsivity, disinhibition, depression, anxiety and apathy. By consequence, it is possible that cognitive and behavioral symptoms in SCA19/22 patients are associated with cerebellar dysfunction. Although the cognitive impairment in the SCA19/22 patients is mild and does not frequently lead to mental retardation, it must not be neglected; it is nevertheless a frequent feature (in 75% of individuals) and has a major impact on quality of life. It is important to note that most of the individuals studied here did not complete their formal education, and three even had slight impairment in receptive language abilities. Given that the symptoms of SCA were combined with neurologic impairments and (in some cases) epilepsy, there is a clear requirement for extended psychosocial care in these patients.

Epilepsy has been described most frequently in SCA10 with variable frequencies among different families. Many of the SCA10 patients have recurrent generalized tonic-clonic seizures, but combinations of myoclonic, complex partial, and secondary generalized tonic-clonic seizures has been described [12]. As in SCA10, three SCA19/22 patients have generalized tonic-clonic seizures or myoclonia. However it seems that the epilepsy in SCA19/22 is not strictly similar to SCA10. To the best of our knowledge, there was no description in SCA10 of benign rolandic epilepsy or generalized atonic seizures. Moreover, in SCA10 patients the epilepsy seems generally to develop a few years after the ataxia. Finally, there is also a clear age effect on the form of the epilepsy, which could also interplay. Potassium channel defects are frequently associated with epilepsy [13], and two SCA19 patients from the original Dutch family had myoclonus [3]. The benign epilepsy phenotypes and EEG of the two French families are similar to those previously reported in one patient with a *de novo KCND3* mutation who had generalized epilepsy with paroxysmal rhythmic theta waves in frontal and parietal regions [14]. Surprisingly, the epilepsy and myoclonus phenotypes recorded to date show male predominance (seven men and only one woman).

These findings suggest that a diagnosis of SCA19/22 should be considered for patients presenting with Parkinsonism, epilepsy, and a family history of neurological disorders. *KCND3* should be included in the list of candidate genes for epilepsy, Parkinsonism and cognitive impairment.

## Conflicts of interest

None.

## Acknowledgement

We thank the patients, their families and the Association “Connaître les Syndromes Cérébelleux” for their participation in this study. We thank Sirine Hamitouche, Alexandre Genet and Kathy Dupont for technical assistance.

## Funding

This work was funded by Lille University Medical Center.

## Author contributions

Vincent Huin was involved in the analysis and interpretation of data (genetic testing) and drafting the manuscript. Isabelle Strubi-Vuillaume was involved in the analysis and interpretation of data (genetic testing). Kathy Dujardin was involved in the collection, analysis and interpretation of data (neuropsychological tests), and drafting and revision of the manuscript. Marine Brion was involved in data collection (neuropsychological tests). Marie Delliaux was involved in data collection (neuropsychological tests). Delphine Dellacherie was involved in the analysis and interpretation of data (neuropsychological tests). Jean-Christophe Cuvellier examined the patients' clinical phenotypes. Luc Defebvre was involved in revising the manuscript. Caroline Moreau examined the patients' clinical phenotypes. Bernard Sablonnière was involved in drafting and revising the manuscript. David Devos examined the patients' clinical phenotypes and was involved in drafting and revising of the manuscript. All authors revised and approved the final manuscript.

**Legends**

**Figure 1:** Pedigrees of the two SCA19/22 families.

Circles denote females, and squares denote males. The arrows indicate the probands. Black filled symbols represent affected individuals. / = deceased. *Genotyped individuals.

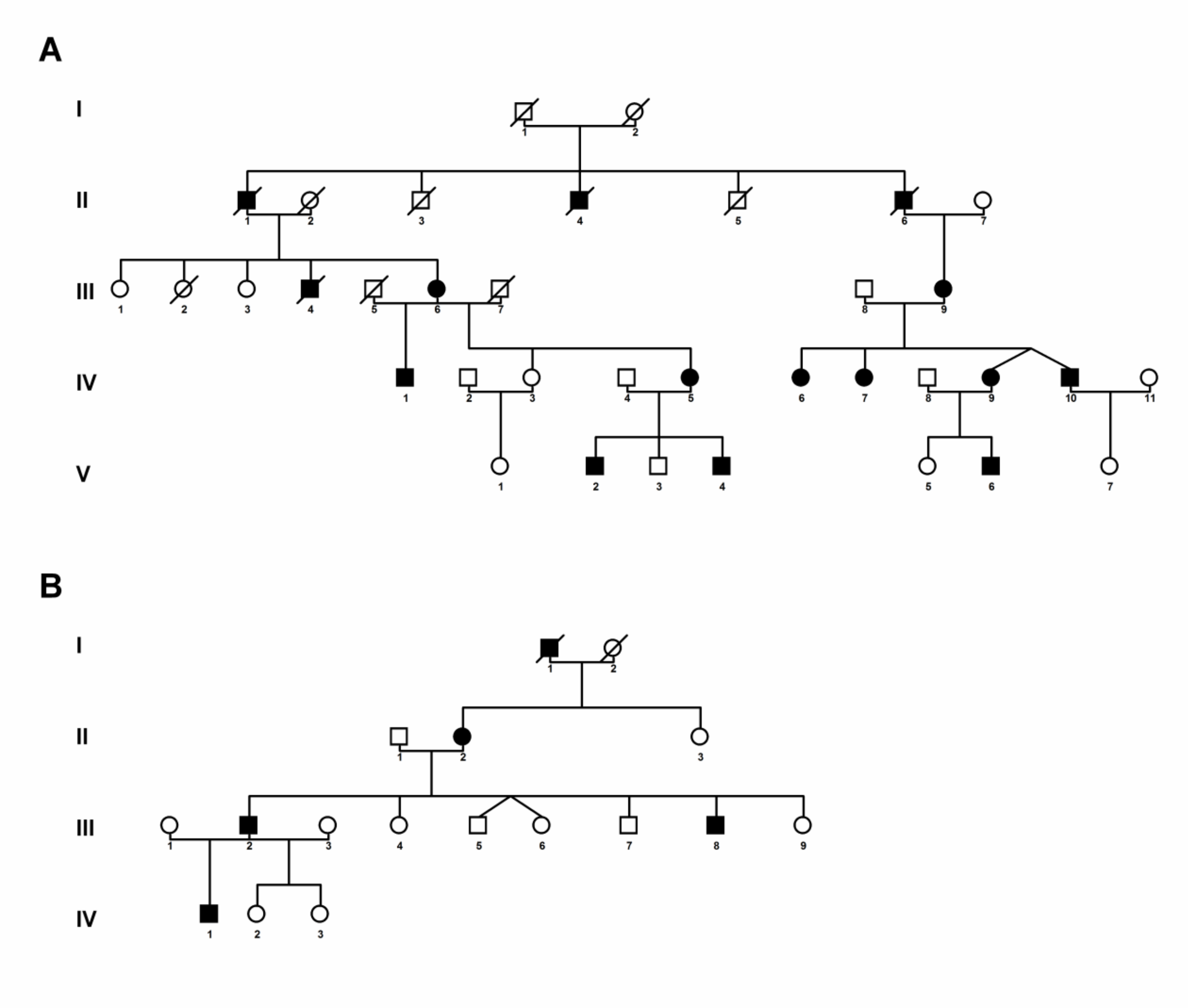

**Table 1:** Clinical features of the 15 affected patients.

**A.** Clinical features. W, working; DW; disabled worker; HM, homemaker; SW, sheltered workplace; UE, unemployed; C, cane; Wa, walker; WC, wheelchair; n.a., not available; NA, not applicable; -, absent; +, mild; ++, moderate, +++, severe; CP, cerebral palsy; De, deafness; N, normal; VA, vermian atrophy. **B.** Cognition, behavior and neuropsychiatric disorders. PS, primary school; SS, secondary school; IQ, intelligence quotient; MoCA: Montreal Cognitive Assessment; BJLO: Benton Judgment of Line Orientation; U, unable to perform the test; *impairment was considered to be significant when performance was 1.5 standard deviations below norms for a similar age range and educational level.

**A CLINICAL FEATURES**

| **Family code** | | A | | | | | | | | | | | B | | | | |
|---|---|---|---|---|---|---|---|---|---|---|---|---|---|---|---|---|---|
| **Subject code** | | III-6 | III-9 | IV-1 | IV-5 | IV-6 | IV-7 | IV-9 | IV-10 | V-2 | V-4 | V-6 | I-1 | II-2 | III-2 | III-8 | IV-1 |
| **Sex** | | Female | Female | Male | Female | Female | Female | Female | Male | Male | Male | Male | Male | Female | Male | Male | Male |
| **Age (year)** | | | | | | | | | | | | | | | | | |
| | current | 80 | 68 | 60 | 50 | 49 | 39 | 33 | 33 | 30 | 15 | 6 | death at 88 | 70 | 52 | 46 | 24 |
| | at onset, ataxia | 66 | 15 | 10 | 35 | 16 | 30 | 23 | 25 | 29 | 4 | - | 20-30 | 20 | 10 | 41 | 2 |
| | at onset, epilepsy | - | - | - | - | - | - | - | 3 | 12 | 3 | 3 | - | - | - | - | n.a. |
| | at first examination | 67 | 53 | 47 | 36 | 36 | 26 | 23 | 3 | 12 | 4 | 3 | 81 | 58 | 41 | 45 | 12 |
| **Impact on quality of life** | | | | | | | | | | | | | | | | | |
| | Working status (age in years when stopped) | W (65) | W (60) | SW (60) | HM | DW (47) | W | W | W | UE | NA | NA | n.a. | W (60) | DW (40) | DW (43) | SW (14) |
| | Mobility aid status (age in years when first used) | - | Wa | - | Wa | - | - | - | - | - | - | - | WC (77) | C (66) | - | - | - |
| **Neurological examination** | | | | | | | | | | | | | | | | | |
| | Gait ataxia | + | ++ | ++ | ++ | + | + | + | + | + | + | + | +++ | ++ | ++ | + | + |
| | Limb ataxia | + | + | + | + | + | + | + | - | - | + | + | ++ | + | + | + | + |
| | Dysgraphia | NA | + | NA | + | + | - | - | - | - | + | + | + | + | + | - | + |
| | Dysarthria | - | + | ++ | + | + | + | - | - | - | - | - | n.a. | + | + | + | - |
| | Dysphagia | - | + | - | - | - | - | - | - | - | - | - | n.a. | + | + | - | - |
| | Oculomotor disturbance | - | - | + | - | - | - | - | - | - | - | + | n.a. | + | + | - | - |
| | Parkinsonism (association of akinesia and rigidity) | + | + | - | + | + | - | + | - | + | - | - | n.a. | + | + | - | - |
| | Epilepsy | - | - | - | - | - | - | - | + | + | + | + | - | - | - | - | + |
| | Pyramidal signs | + | - | - | + | - | - | - | - | - | - | - | + | - | + | - | - |
| **Other clinical features** | | - | - | - | - | - | - | - | - | - | CP | - | - | - | De | - | - |
| **Brain MRI** | | n.a. | VA | VA | n.a. | VA | VA | n.a. | n.a. | N | n.a. | N | n.a. | VA | n.a. | n.a. | n.a. |

**B COGNITION, BEHAVIOR AND NEUROPSYCHIATRIC DISORDERS**

| | | | | | | | | | | | | | | | | | |
|---|---|---|---|---|---|---|---|---|---|---|---|---|---|---|---|---|---|
| **Overall cognition** | | | | | | | | | | | | | | | | | |
| | Cognitive impairment | ++ | ++ | ++ | ++ | - | - | + | + | + | + | + | n.a. | - | + | + | + |
| | Education | PS | PS | PS | SS | SS | n.a. | SS | PS | SS | NA | NA | n.a. | PS | PS | SS | PS |
| | Time in formal education (years) | 8 | 8 | n.a. | 10 | 14 | n.a. | 12 | 12 | 11 | NA | NA | n.a. | 8 | 10 | 11 | 10 |
| | Special education | - | - | + | - | - | n.a. | - | + | + | + | NA | n.a. | - | - | - | + |
| | Learning delay | + | + | ++ | - | - | n.a. | - | + | + | + | + | n.a. | + | + | - | ++ |
| | IQ (PM47) | <70 | >80 | 80 | 60-70 | >80 | n.a. | >80 | >80 | >80 | n.a. | n.a. | n.a. | >80 | >80 | >80 | > 80 |
| | MoCA (/30) | 10 | 20 | 17 | 18 | 28 | n.a. | 25 | 22 | 24 | n.a. | n.a. | n.a. | 28 | 22 | 24 | 23 |
| **Attention and working memory** | | | | | | | | | | | | | | | | | |
| | Corsi forward span | 4 | 4 | 4 | 5 | 4 | n.a. | 5 | 4 | 4 | n.a. | n.a. | n.a. | 5 | 5 | 6 | 6 |
| | Corsi backward span | 3 | 4 | 3 | 4 | 5 | n.a. | 5 | 5 | 5 | n.a. | n.a. | n.a. | 4 | 5 | 6 | 5 |
| | Simple reaction time (ms) | 749 | 364 | 378.5 | 247 | 324.5 | n.a. | 233.5 | 315.5 | 274.5 | n.a. | n.a. | n.a. | 344.5 | 303 | 234 | 285 |
| | Go No Go (ms) | 657 | 506 | 429 | 594 | 531.5 | n.a. | 355.5 | 413 | 424 | n.a. | n.a. | n.a. | 563.5 | 404.5 | 356.5 | 442.5 |
| | Divided attention (ms) | 824.5 | 727 | 502 | 622 | 730 | n.a. | 464 | 497.5 | 486 | n.a. | n.a. | n.a. | 752 | 501 | 473.5 | 577.5 |
| **Visuospatial learning and memory** | | | | | | | | | | | | | | | | | |
| | 10/36 immediate recall (/30) | 13 | 14 | 11 | 12 | 13 | n.a. | 22 | 14 | 17 | n.a. | n.a. | n.a. | 9 | 20 | 16 | 19 |
| | 10/36 delayed recall (/10) | 4 | 4 | 4 | 4 | 4 | n.a. | 8 | 4 | 4 | n.a. | n.a. | n.a. | 4 | 7 | 5 | 6 |
| **Visuospatial function** | | | | | | | | | | | | | | | | | |
| | BJLO (/15) | 3 | 9 | 4 | 6 | 14 | n.a. | 13 | 15 | 10 | n.a. | n.a. | n.a. | 11 | 13 | 6 | 12 |
| | BJLO (/30) | 14 | 22 | 15 | 19 | 29 | n.a. | 28 | 30 | 24 | n.a. | n.a. | n.a. | 26 | 28 | 19 | 27 |
| **Language** | | | | | | | | | | | | | | | | | |
| | Token test (/36) | 23.5 | 29 | 33 | 30 | 34 | n.a. | 34 | 34 | 33 | n.a. | n.a. | n.a. | 32.5 | 34 | 33 | 33 |
| **Executive Functions** | | | | | | | | | | | | | | | | | |
| Tower of London | | | | | | | | | | | | | | | | | |
| | 3N items | 3.33 | 3.67 | U | 3.33 | 3 | n.a. | 3 | 3 | 3 | n.a. | n.a. | n.a. | 10.33 | 3 | 3.33 | 3.67 |
| | 5N items | 11 | 8 | U | 6.33 | 8.67 | n.a. | 5 | 9.67 | 11.67 | n.a. | n.a. | n.a. | 6.67 | 5.67 | 6.67 | 9.33 |
| | 5I+ items | 5 | 5.33 | U | 5 | 5 | n.a. | 7.67 | 8.67 | 5 | n.a. | n.a. | n.a. | 6.33 | 5 | 6.33 | 7.67 |
| | 5I- items | 12 | 8.67 | U | 9.67 | 7 | n.a. | 8.33 | 8.67 | 7.67 | n.a. | n.a. | n.a. | 7.67 | 9.33 | 9 | 6.33 |

**Video 1:** Video of patient IV-6 in family A, mildly affected by a cerebellar syndrome.

Video 1 Patient IV-6.m4v

**Video 2:** Video of patient III-9 in family A, mother of patient IV-6 and more severely affected by a cerebellar syndrome.

Video 2 Patient III-9.m4v

**HIGHLIGHTS**

- SCA19 and SCA22 (caused by *KCND3* gene mutations) are rare forms of inherited ataxia.
- We observed Parkinsonism is a high proportion of individuals with SCA19/22.
- Epilepsy seems to be a feature of SCA19/22.
- We characterized cognitive and behavioral disorders.
- *KCND3* may be a candidate gene for epilepsy, Parkinsonism and cognitive disorders.

**SUPPORTING INFORMATION**

**Data S1: Next-generation sequencing**

**Gene panel design:** A list of the 24 genes associated with dominant hereditary ataxia was selected and included in an SCA panel after analyzing the literature and the OMIM database. Genes with trinucleotide repeat expansions and those associated with SCA were not included in the panel. The gene panel comprised *AFG3L2* (SCA28), *ATP1A2* (episodic ataxia), *ATP1A3* (episodic ataxia), *CACNA1A* (episodic ataxia type 2), *CACNB4* (episodic ataxia type 5), *CCDC88C* (SCA40), *EEF2* (SCA26), *ELOVL4* (SCA34), *ELOVL5* (SCA38), *FGF14* (SCA27), *IFRD1* (SCA18), *ITPR1* (SCA15/16/29), *KCNA1* (episodic ataxia type 1), *KCNC3* (SCA13), *KCND3* (SCA19/22), *PDYN* (SCA23), *PLEKHG4* (SCA4), *PRKCG* (SCA14), *SLC1A3* (episodic ataxia type 6), *SL2A2* (episodic ataxia), *SPTBN2* (SCA5), *TGM6* (SCA35), *TTBK2* (SCA11), and *TMEM240* (SCA21). The SCA panel of 24 genes was designed using SureDesign software (v4.5, Agilent Technologies, Santa Clara, CA, USA).

**Targeted sequencing:** Exon capture of the 24 selected genes and libraries was prepared using the Haloplex kit (Agilent Technologies), according to the manufacturer's instructions. Pooled libraries (n=20) prepared using the SCA panel were sequenced on a MiSeq system (Illumina) with MiSeq Reageant kit v2.

**Bioinformatics analysis:** Sequence analysis was performed using SeqPilot (v4.3.1, JSI Medical Systems, Ettenheim, GermanyJSI) and MiSeqReporter (Illumina) software (v2.6, Illumina, San Diego, CA, USA) programs. Variants were annotated and filtered with Variant Studio software (v3.0, Illumina, San Diego, CA, USA). Harmful effects were predicted with Alamut software (v2.0, Interactive BioSoftware, Rouen, France).

**Supplemental Figure S1:** Electrophoregrams of the heterozygous *KCND3* mutation. Electrophoregrams obtained by sequencing exon 2 of *KCND3* showed a normal sequence in the unaffected individual (upper panel) and the c.679_681delTTC p.F227del mutation in a patient (lower panel).

***KCND3*** **c.679_681delTTC, p.F227del**

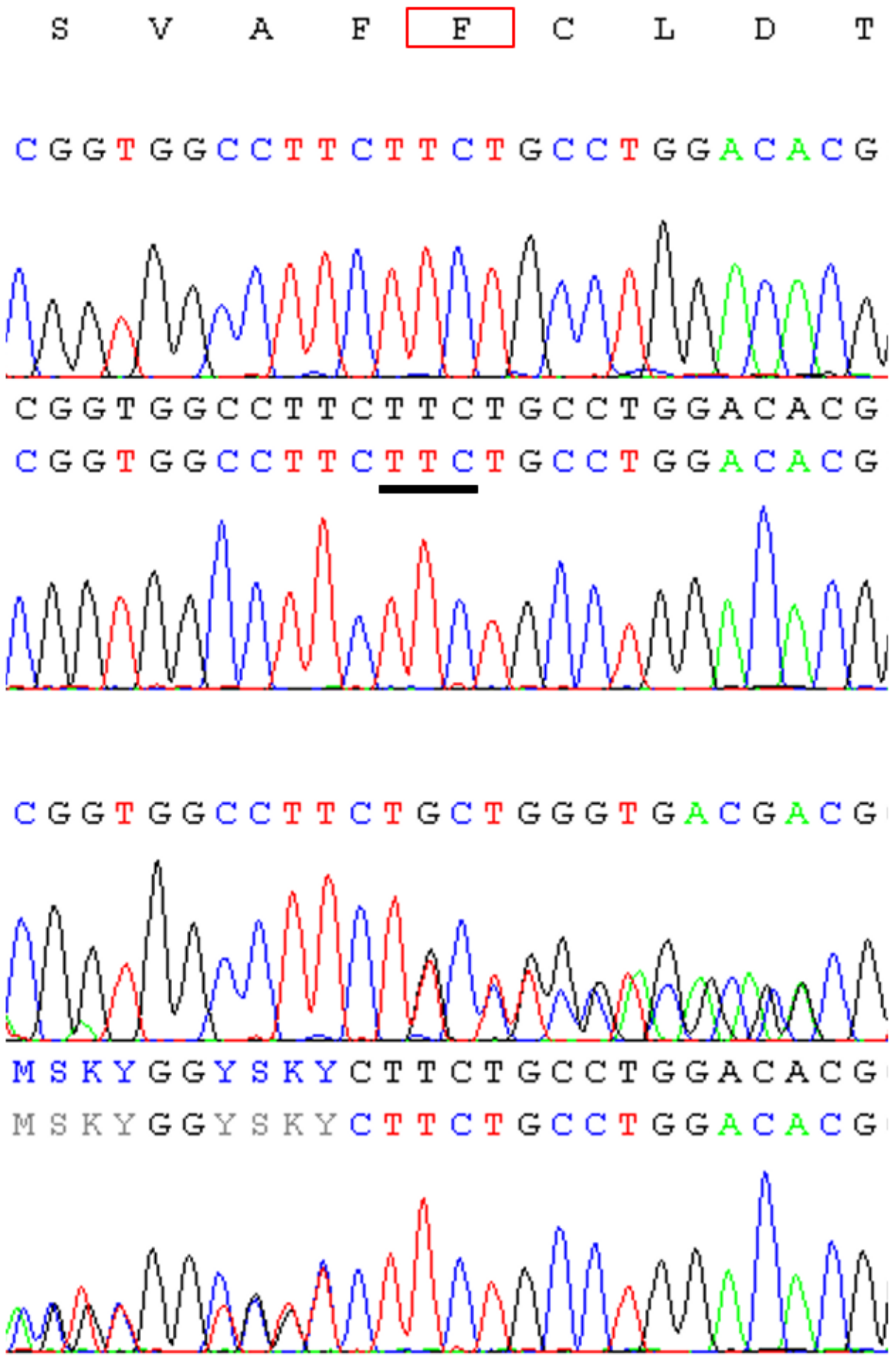

**Supplemental Figure S2:** Electroencephalograms of two SCA19/22 patients.

**A.** EEG of patient V-4 in family A revealed short bursts of pointed theta waves in the right frontal region. **B.** EEG of the patient V-6 in family A showing abundant bursts of bilateral-synchronous spike and waves discharges which are more prevalent in the right frontal region.

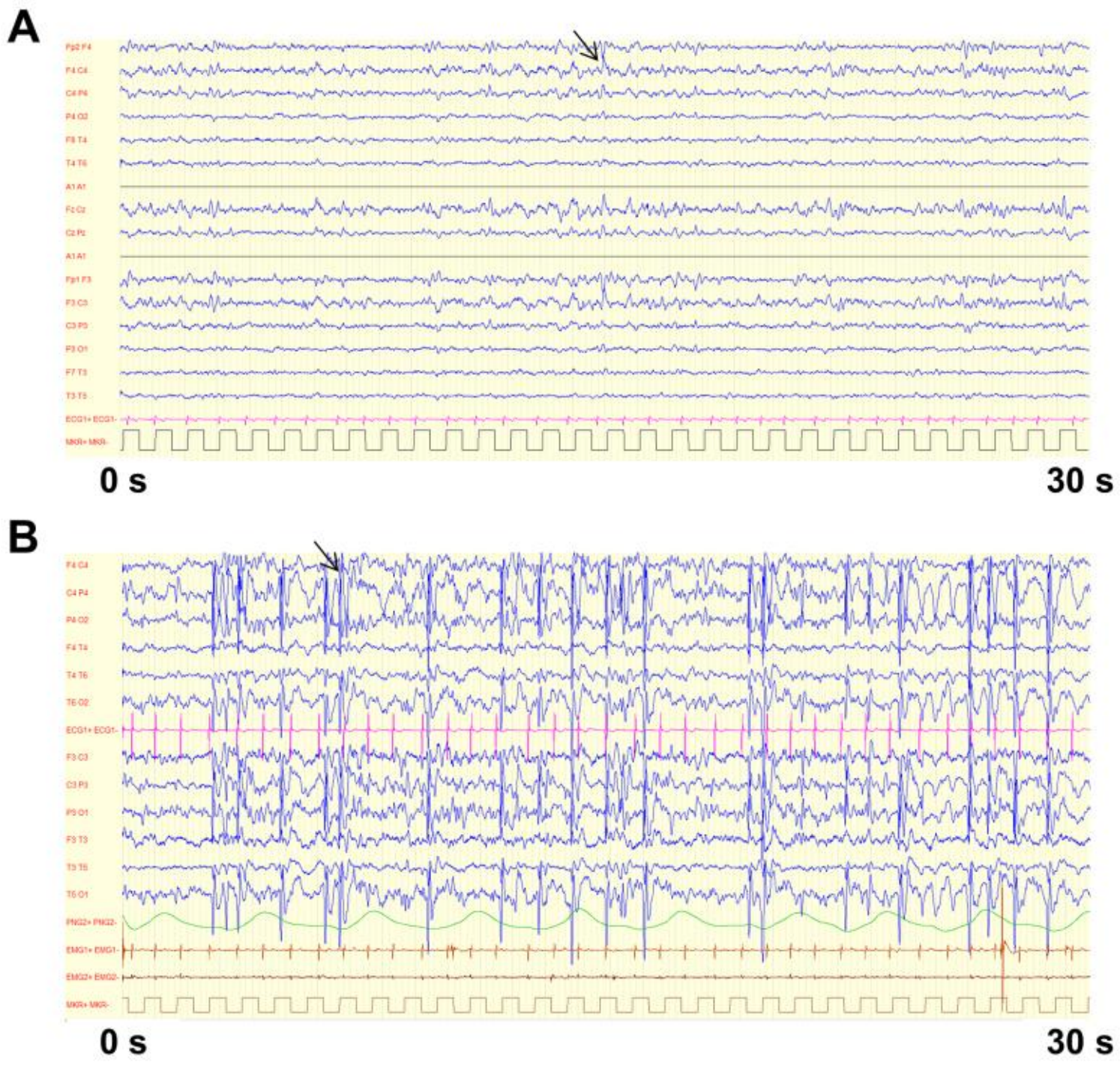